\documentclass
[superscriptaddress,secnumarabic,amssymb,amsmath,nobibnotes,aps,prd,showkeys,showpacs,nofootinbib,onecolumn]{revtex4}%
\usepackage{graphicx}
\usepackage{epsf}
\usepackage{bm}
\usepackage{amsmath}
\usepackage{amsfonts}
\usepackage{amssymb}
\usepackage{epstopdf}%
\providecommand{\U}[1]{\protect\rule{.1in}{.1in}}

\newcommand{\be}{\begin{equation}}
\newcommand{\ee}{\end{equation}}
\newcommand{\bear}{\begin{eqnarray}}
\newcommand{\eear}{\end{eqnarray}}

\newcommand{\mincir}{\raise
-3.truept\hbox{\rlap{\hbox{$\sim$}}\raise4.truept\hbox{$<$}\ }}
\newcommand{\magcir}{\raise
-3.truept\hbox{\rlap{\hbox{$\sim$}}\raise4.truept\hbox{$>$}\ }}

\begin{document}
\title{Finsler-Randers Cosmology: dynamical analysis and growth of matter perturbations}
\author{G. Papagiannopoulos}
\email{yiannis.papayiannopoulos@gmail.com}
\affiliation{Faculty of Physics, Department of Astronomy-Astrophysics-Mechanics University
of Athens, Panepistemiopolis, Athens 157 83, Greece}
\author{S. Basilakos}
\email{svasil@academyofathens.gr}
\affiliation{Academy of Athens, Research Center for Astronomy and Applied Mathematics,
Soranou Efesiou 4, 11527, Athens, Greece}
\author{A. Paliathanasis}
\email{anpaliat@phys.uoa.gr}
\affiliation{Instituto de Ciencias F\'{\i}sicas y Matem\'{a}ticas, Universidad Austral de
Chile, Valdivia, Chile}
\affiliation{Institute of Systems Science, Durban University of Technology, PO Box 1334,
Durban 4000, Republic of South Africa}
\author{S. Savvidou}
\email{sophie.savvidou@gmail.com}
\affiliation{Faculty of Physics, Department of Astrophysics - Astronomy - Mechanics
University of Athens, Panepistemiopolis, Athens 157 83, Greece}
\author{P.C. Stavrinos}
\email{pstavrin@math.uoa.gr}
\affiliation{Department of Mathematics, National \& Kapodistrian University of Athens,
Athens 15784, Greece}

\begin{abstract}
We study for the first time the dynamical properties and the growth index of
linear matter perturbations of the Finsler-Randers (FR) cosmological model,
for which we consider that the cosmic fluid contains matter, radiation and a
scalar field. Initially, for various FR scenarios we implement a critical
point analysis and we find solutions which provide cosmic acceleration and
under certain circumstances we can have de-Sitter points as stable late-time
attractors. Then we derive the growth index of matter fluctuations in various
Finsler-Randers cosmologies. Considering cold dark matter and neglecting the
scalar field component from the perturbation analysis we find that the
asymptotic value of the growth index is $\gamma_{\infty}^{(FR)}\approx\frac
{9}{16}$, which is close to that of the concordance $\Lambda$ cosmology,
$\gamma^{(\Lambda)} \approx\frac{6}{11}$. In this context, we show that the
current FR model provides the same Hubble expansion with that of Dvali,
Gabadadze and Porrati (DGP) gravity model. However, the two models can be
distinguished at the perturbation level since the growth index of FR model is
$\sim18.2\%$ lower than that of the DPG gravity $\gamma^{(DGP)} \approx
\frac{11}{16}$. If we allow pressure in the matter fluid then we obtain
$\gamma_{\infty}^{(FR)}\approx\frac{9(1+w_{m})(1+2w_{m})}{2[8+3w_{m}%
(5+3w_{m})]}$, where $w_{m}$ is the matter equation of state parameter.
Finally, we extend the growth index analysis by using the scalar field and we
find that the evolution of the growth index in FR cosmologies is affected by
the presence of scalar field.


\end{abstract}
\maketitle
\date{\today}

\section{\bigskip Introduction}

{\label{intro} During the last decade the interest of many researchers
of the scientific community has been increased to applications of Finsler
geometry for gravitation and cosmology
\cite{Rad41,Stav07,2,3,4,5,6,7,Bastav13,Bas2013,10,11}. From the geometrical
view point the relativistic extensions of Finsler geometry have received a lot
of attention, due to the fact that this geometry smoothly extends the nominal
Riemann geometry \cite{Rad41}. Of course the Riemannian geometry can be seen
as a special case of the Finslerian one. According to these lines, Finsler
gravity is considered as the simplest family of generalizations, because it
naturally extends general relativity. Generally, one needs to start with the
Lorentz symmetry breaking, which is a common property within quantum gravity
phenomenology. Such a departure from relativistic symmetries of space-time,
leads to the possibility for the underlying physical manifold to have a richer
geometric structure than the simple pseudo-Riemann geometry. Therefore, it has
been proposed that Finsler gravity can be used towards studying the physical
phenomena in the universe, among which the implications of quantum gravity and
the related Lorentz violations in the early universe and the accelerated
expansion of the universe prior to the present epoch.}



The development of research in modified gravity theories for studying
universe's evolution can be combined with a locally anisotropic structure of
the Finslerian gravitational field\cite{14,Stavv2014}. \newline The Finslerian
metric structure contains coordinates of position-x of a base manifold M and
velocity or direction y-coordinates $y=\frac{dx}{dt}$ on the tangent
(internal) space of $M$. In this case the intrinsic dynamical form of Finsler
geometry extends the limits of Riemannian framework as well as of
gravitational field because of internal variables. The y-dependence
essentially characterizes the Finslerian gravitational field and has been
combined with the concept of anisotropy which causes the deviation from
Riemannian geometry. Therefore the consideration of Finsler geometry as a
candidate for studying of gravitational theories provides that matter dynamics
take place. In the theory of Finslerian gravitational field a peculiar
velocity field is produced by the gravity of mass fluctuations which are due
to the anisotropic distribution and motion of particles \cite{Stav07,12}. On
the other hand the wide metric structure of Finsler geometry with torsions,
more than one covariant derivatives and anisotropic curvatures extends the
framework of field equations in the general relativity and cosmology. A
unified description of the Finslerian gravitational field of a spacetime
manifold $M$ is given by a metric function F, a total metric structure g on
the tangent bundle of $M$, a metrical compatible connection and a non linear
connection $N$ \cite{14,13}. The Cartan's type torsion tensor characterizes
all the geometrical concepts of Finsler geometry and contributes in its
configuration as a physical geometry \cite{20} . Finsler - Cartan
gravitational field theory is compatible with the structure of general
relativity. However in many cases we restrict our consideration in order to
describe the local anisotropic ansantz of gravitational field equations on a
four-dimensional Finsler manifold \cite{Stav07}.

In the context of studying of Finsler cosmology it has been considered the
form of Riemannian \emph{osculation} of a metric \cite{15}. The Finslerian
metric tensor and a contravariant vector field $y^{i}(x)$ may be used to
construct the Riemannian metric tensor $a_{ij}(x)=g_{ij}(x,y(x))$ \ in a
domain of space-time. The Riemannian space associated with this metric tensor
is called osculating Riemannian space. This gives us the possibility to view
cosmological considerations in four dimensional space time framework. Models
have been build to study gravitational theories that are constructed in
Finsler space time. A special type of Finsler space is the Finsler - Randers
space hereafter (FR) \cite{16} which constitutes an important geometrical
structure in Finsler spaces \cite{17} as far as its applications. In the
general relativity and cosmology are concerned
eg.\cite{Stav07,Bastav13,10,14,18}.This type of space expresses a locally
anisotropic perturbation of Riemannian geometry. The FR cosmological model was
introduced in \cite{19}, \cite{Stav07} and has been studied further in the
framework of general relativity and cosmology.

Some considerations concerning the geodesics in Finsler geometries are in
order at this point. In a Finsler spacetime the geodesics include anisotropic
terms due to the presence of the Cartan tensor which affects all the
geometrical concepts of spacetime. In Finsler-Randers space particularly the
form of geodesics is given in Ref.\cite{18}. It is obvious that the extra
terms of geodesics are introduced by the co-vector $u_{\mu}$ (see next
section) { of the second term of the Finsler-Randers metric which gives a
rotation in the geodesic equation (see Stavrinos et al. \cite{Rad41}). From
the physical viewpoint an observer along the geodesics in a Finsler-Randers
spacetime is rotating giving an extension of the geodesic equation of a
pseudo-Riemannian spacetime. Consequently, the geodesics in Finsler-Randers
space are influenced by Lorentz violation phenomena (because of the vector
}$u_{\mu}${) which are related to redshift, luminosity distance and dispersion
relation. In a Finslerian gravitational field where an intrinsic anisotropy
takes part the ticking rate of clocks is influenced from the orientation of
the universal gravitational field which produces a different frequency of
light (see Ref.\cite{Answ1}). Recently, Hohmann \& Pfeifer \cite{7} discussed
this issue in a very pedagogical way by studying the magnitude-redshift
relation and the deceleration parameter in the Finsler cosmological
background, including that of Finsler-Randers model. The results of Hohmann \&
Pfeifer \cite{7} allow a confrontation of these geometries with supernovae
type I data.}

In previous publications some of us \cite{Bastav13} discussed the
observational consequences of flat FR model and demonstrated the compatibility
of this scenario with the braneworld Dvali, Gabadadze and Porrati (hereafter
DGP; \cite{DGP}) model. As found in \cite{Bastav13} the two cosmological
models are cosmologically identical, despite the fact that they constructed
from different background geometries. On the other hand, the fact that DGP
gravity is under observational pressure \cite{Fair06,Maa06,Song07} implies
that the flat FR model suffers from the same problem \cite{Bastav13}.
{The latter issue has led Basilakos et al. \cite{Bas2013} to propose an
extended version of the FR model based on geometrical arguments, free from
observational inconsistencies. In particular, Basilakos et al. \cite{Bas2013}
found that the geometrical extension of FR model affects the Hubble expansion
via the function }$\Psi(a)$\footnote{The Hubble parameter in the extended
Finsler-Randers model is given by $H^{2}(a)=\frac{8\pi G}{3}\rho_{m}%
-a^{-2}\int a\Psi(a) da-{\frac{C_{1}}{a^{2}}}$ [see Eqs.(32) and (39) in
Ref.\cite{Bas2013}].}{, which is defined on a geometrical basis. Of course as
clearly explained in Ref.\cite{Bas2013}, the fact that the precise functional
form of the }$\Psi(a)${ parameter can not be found from first principles
implies that the only way to use the extended Finsler Randers approach in
cosmology is to phenomenologically select the functional form of }$\Psi(a)${
using some well known dark energy models as reference models. Therefore, the
geometrical structure of }$\Psi(a)${ in Ref.\cite{Bas2013} is indirectly
related with the scalar field. Instead of doing that in the current paper we
have asked ourselves under which conditions the standard Finsler-Randers
approach provides cosmological models which can be consistent with
observations, namely to explain the accelerated expansion of the universe. In
this work, we propose a scalar field description of the standard
Finsler-Randers model by introducing a scalar field directly in the field
equations. Notice, that this general path has been used extensively in the
literature for other cosmological models (see for example the scalar tensor
theories e.g. Brans-Dicke). Of course, in our case the precise form of the
potential }$V(\phi)${ is still unknown which however is also the case for the
scalar field dark energy models which adhere to General Relativity and for the
extended Finsler-Randers model (see Ref.\cite{Bas2013}) as far as the
functional form of }$\Psi(a)${ is concerned. Clearly, each approach has
advantages and disadvantages, therefore we believe that it is interesting to
investigate the possibility to have a scalar field in the context of
Finsler-Randers gravity. Furthermore, we would like to stress that it is
traditional to study, in each proposed model, its critical points, in order to
have an impression of the dynamical consequences of the model. Such an
analysis has not been done for the Finsler-Randers model and we believe that
the present study covers this gap in the literature.}

In this article, using a standard scalar field language we thoroughly
investigate the main properties of the FR model at the cosmological and
perturbation levels respectively and be seen as a natural extension of the
previously published works \cite{Bastav13,Bas2013}. Specifically, the lay out
the manuscript is as follows. In section II we present the main features of
the FR cosmological model in a FLRW metric and provide the field equations in
the scalar field approach. In section III we perform a dynamical analysis by
studying the critical points of the field equations in the dimensionless
variables for two type of potentials (exponential and hyperbolic potentials)
and we compare our results with those of General Relativity (GR). We find that
there are common critical points between FR and GR but the stability changes,
while some new critical points appear in FR gravity. As an example, when the
potential is exponential we find a unique stable point that describes a de
Sitter universe. In section IV we explore the behavior of the FR model at the
perturbation level and we compare its predictions regarding the growth index
of linear matter perturbations with those of DGP and $\Lambda$CDM models
respectively. Finally, we present our conclusions in Section V.

\section{Finsler-Randers theory}

\label{section2} The geometrical origin of Finsler-Randers (FR) cosmological
model is based on the Finslerian geometry which is a natural generalization of
the standard Riemannian geometry. In this section we present only the main
ingredients of the FR model, for more details we refer the reader to
Refs.\cite{15,Bekenstein,Mir,Bao,13}). In general, having a manifold $M$, a
Fisnler space is produced from a generating differentiable function $F(x,y)$
on a tangent bundle $F:TM\rightarrow R,TM=\tilde{T}(M)\backslash\{0\}$, where
$F$ is a one degree homogeneous function and the variable $y$ is related with
the time derivative of $x$, $y=\frac{dx}{dt}$. In this context, the FR space
time is a special case in which the aforementioned metric function is given
by
\[
F(x,y)=\ \sigma(x,y)+u_{\mu}(x)y^{\mu},\;\;\;\sigma(x,y)=\sqrt{a_{\mu\nu
}y^{\mu}y^{\nu}},
\]
where $u_{\mu}=(u_{0},0,0,0)$ is a weak primordial vector field with $\Vert
u_{\mu}\Vert\ll1$ and $a_{\mu\nu}$ is the metric of the Riemannian space. The
Finslerian contribution is provided by the vector field $u_{\mu}$ which
introduces a preferred direction in space time. As well the field $u_{\mu}$
causes a differentiation of geodesics from a Riemannian spacetime \cite{18}.
Using the Hessian of $F$ we can write the Finslerian metric tensor
\[
f_{\mu\nu}=\ \frac{1}{2}\frac{\partial^{2}F^{2}}{\partial y^{\mu}\partial
y^{\nu}},
\]
from which we derive the Cartan tensor $C_{\mu\nu k}=\frac{1}{2}\frac{\partial
f_{\mu\nu}}{\partial y^{k}}$. It is interesting to mention that the component
$u_{0}$ is given in terms of $C_{000}$ via $u_{0}=2C_{000}$ \cite{Stav07}.

Now, the field equations in the FR\ cosmology are written as
\begin{equation}
L_{\mu\nu}=\ 8\pi G(T_{\mu\nu}-\frac{1}{2}Tg_{\mu\nu}),\label{EE}%
\end{equation}
\bigskip where $L_{\mu\nu}$ is the Finslerian Ricci Tensor {(for more
details see the book of Asanov \cite{Rad41})}, $g_{\mu\nu}=\frac{Fa_{\mu\nu}%
}{\sigma}$, $~T_{\mu\nu}$ is the energy momentum tensor and $T$ its trace.
{In a forthcoming paper we attempt to obtain the field equations of
Finsler gravity using a Lagrangian formalism (Triantafilopoulos et al. in
preparation). Such an analysis will help towards understanding the
geometrical/dynamical properties of Finsler gravity.}

Now, if we model the expanding universe as a Finslerian perfect fluid with
velocity 4-vector field $u_{\mu}$ then the energy momentum tensor is written
as $T_{\mu\nu}=\mathrm{diag}\left(  \rho,-Pf_{ij}\right)  $, where $\{\mu
,\nu\}\in\{0,1,2,3\}$ and $\{i,j\}\in\{1,2,3\}$.
Here $\rho$ is the total energy density and $p$ is the corresponding pressure.
Using the latter form of energy-momentum tensor and the spatially flat FLRW
metric\footnote{For the benefit of the reader we provide the nonzero
components of the Ricci tensor in the context of Finsler-Randers geometry.
These are: $L_{00}=3(\frac{\ddot{a}}{a}+3\frac{\dot{a}}{4a}\dot{u}_{0})$ and
$L_{ii}=-(a\ddot{a}+2\dot{a}^{2}+\frac{11}{4}a\dot{a}\dot{u}_{0})/\Delta_{ii}$
where $(\Delta_{11},\Delta_{22},\Delta_{33})=(1,r^{2},r^{2}\mathrm{sin}%
^{2}\theta)$.},
\[
ds^{2}=-dt^{2}+a^{2}\left(  t\right)  \left(  dx^{2}+dy^{2}+dz^{2}\right)  ,
\]
the gravitational field equations provide the modified Friedmann's Equations
\cite{Stav07}
\begin{equation}
\dot{H}+H^{2}+\frac{3}{4}HZ_{t}=-\frac{4\pi G}{3}(\rho+3P), \label{e1}%
\end{equation}%
\begin{equation}
\dot{H}+3H^{2}+\frac{11}{4}HZ_{t}=4\pi G(\rho-P), \label{e2}%
\end{equation}
where the over-dot represents derivative with respect to the cosmic time $t$,
$H={\dot{a}}/a$ is the Hubble parameter
and $Z_{t}=\dot{u}_{0}<0$ (see Ref. \cite{Stav07}).
Combining Eqs. (\ref{e1}) and (\ref{e2}) we arrive at
\begin{equation}
H^{2}+HZ_{t}=\frac{8\pi G}{3}\rho
\label{6}
\end{equation}
or
\begin{equation}
H^{2}=\left(  \sqrt{\frac{8\pi G}{3}\rho+\frac{Z_{t}^{2}}{4}}-\frac{Z_{t}}%
{2}\right)  ^{2}. \label{66}%
\end{equation}
Moreover, the Bianchi identities which insures the covariance of the theory
imposes the following conservation equation:
\begin{equation}
\dot{\rho}+3H\left(  \rho+p\right)  =0. \label{e3}%
\end{equation}
Clearly, the dynamics of the universe is affected by the the extra term
$H(t)Z_{t}$. As expected, in the case of $u_{0}\equiv0$ (or $C_{000}\equiv0$,
$F/\sigma=1$), which means $Z_{t}=0$, the modified Friedmann equation
(\ref{6}) takes the nominal form.

Up to this point, we did not specify the physics of the cosmic fluids
involved. Let us assume that we have a mixture of two fluids, matter and
scalar field, hence the total density and pressure are given by
\begin{equation}
\rho=\rho_{m}+\rho_{\phi},\;\;\;\;\;p=p_{m}+p_{\phi}. \label{Eq1}%
\end{equation}
Notice, that $\rho_{m}$ is the matter density, $\rho_{\phi}$ is the density of
the scalar field and $(p_{m},p_{\phi})$ are the corresponding pressures. In
this work, we restrict our analysis in the presence of barotropic cosmic
fluids, where the corresponding equation of state (EoS) parameters are given
by $w_{m}=p_{m}/\rho_{m}$ and $w_{\phi}=p_{\phi}/\rho_{\phi}$. In the rest of
the paper we assume of constant $w_{m}$ (cold $w_{m}=0$ and relativistic
$w_{m}=1/3$ matter), while the EoS parameter $w_{\phi}$ is given in terms of
the scalar field. Indeed using a standard scalar field language one can write
$\rho_{\phi}$ and $p_{\phi}$ as follows
\begin{equation}
\rho_{\phi}=\frac{1}{2}\dot{\phi}^{2}+V\left(  \phi\right)  ~,~p_{\phi}%
=\frac{1}{2}\dot{\phi}^{2}-V\left(  \phi\right)  , \label{Eq4}%
\end{equation}
and thus
\[
w_{\phi}=\frac{({\dot{\phi}}^{2}/2)-V(\phi)}{({\dot{\phi}}^{2}/2)+V(\phi)}\;,
\]
where $V(\phi)$ is the potential energy of $\phi$ (for review see \cite{Ame10}).

Under the above conditions, the conservation law (\ref{e3}) becomes
\begin{equation}
\label{Conv}{\dot\rho}_{m}+ 3 H(1+w_{m})\rho_{m} + {\dot\rho}_{\phi}+ 3
H(1+w_{\phi})\rho_{\phi}= 0
\end{equation}
or equivalently
\begin{align}
\dot{\rho}_{m}+3 H(1+w_{m})\rho_{m}  &  =Q,\label{Eq2}\\
\dot{\rho}_{\phi}+3H(1+w_{\phi})\rho_{\phi}  &  =-Q, \label{Eq3}%
\end{align}
where $Q$ is the rate of interaction between the scalar field and the matter
source.
Below, we briefly present two forms of the interaction rate.

\subsection{Finsler-Randers with minimally coupled fluids}

The absence of interaction between matter and scalar field implies $Q=0$ which
means that the two cosmic fluids are minimally coupled. Within this framework,
solving Eqs.(\ref{Eq2}) and (\ref{Eq3}) we arrive at
\begin{equation}
\rho_{m}=\rho_{m0}a^{-3(1+w_{m})},\;\;\;\;\;\;\rho_{\phi}=\rho_{\phi0}X(a)
\label{frie55}%
\end{equation}
with
\begin{equation}
\label{frie555}X(a)={\exp}\left(  3\int_{a}^{1} [1+w_{\phi}(a)]d\mathrm{ln}%
a\right)  ,
\end{equation}
where $\rho_{m 0}$ and $\rho_{\phi0}$ are the matter and scalar field
densities at the present time. Therefore, with the aid of (\ref{frie55})
equation (\ref{66}) becomes
\begin{align}
\label{EE11}E^{2}(a)=\left[  \sqrt{\Omega_{Z_{t}}+\Omega_{m0}a^{-3(1+w_{m})}+
\Omega_{\phi0}X(a)}+\sqrt{\Omega_{Z_{t}}} \right]  ^{2}%
\end{align}
where $E(a)=H(a)/H_{0}$, $\sqrt{\Omega_{Z_{t}}}=-\frac{Z_{t}}{2H_{0}}$ $H_{0}$
is the Hubble constant, $\Omega_{m0}=8\pi G\rho_{m0}/3H_{0}^{2}$ and
$\Omega_{\phi0}=8\pi G\rho_{\phi0}/3H_{0}^{2}$. Of course it is trivial to
show that for a constant EoS parameter we have $X(a)=a^{-3(1+w_{\phi})}$.

Inserting the condition $E(1)=1$ into Eq.(\ref{EE11}) and after some simple
calculations we find
\begin{equation}
\label{nfe22}E^{2}(a)= \Omega_{m0}a^{-3(1+w_{m})}+\Omega_{\phi0}X(a)
+2\Omega_{Z_{t}}+2\sqrt{\Omega_{Z_{t}}} \sqrt{\Omega_{m0}a^{-3(1+w_{m}%
)}+\Omega_{\phi0}X(a)+\Omega_{Z_{t}}}%
\end{equation}
where $\Omega_{Z_{t}}=(1-\Omega_{m0}-\Omega_{\phi0})^{2}/4$.

We immediately recognize the following situation. In the case of $w_{m}=0$
(non-relativistic matter) and $\rho_{\phi}=0$ (or $\Omega_{\phi0}=0$) we
recover the standard spatially flat FR model which has been studied in several
papers (see \cite{Rad41,Stavrinos2005,Stav07,Kouretsis:2012ys}. Interestingly,
Basilakos \& Stavrinos \cite{Bastav13} have shown that the latter FR scenario
is equivalent with the spatially flat Dvali, Gabadadze and Porrati (DGP)
cosmological model as far as the Hubble parameter is concerned. As discussed
in \cite{Bastav13} the cosmological data, especially those of baryon acoustic
oscillation (BAO), cosmic microwave background (CMB) shift parameter
\cite{Maa06} and integrated Sachs-Wolfe (ISW) effect \cite{Song07}, disfavor
either DGP or FR gravity models respectively. In order to overcome the above
observational problem, Basilakos et al. \cite{Bas2013} proposed an extended
form of Finsler-Randers model and they proved that it can resemble a large
family of non-interacting dark energy scenarios at the background level. We
would like to stress that the setup of Basilakos et al. \cite{Bas2013} was
based only on geometrical arguments, while the current version of the extended
FR gravity model is within the framework of scalar field theory.

\subsection{Finsler-Randers versus matter/scalar field interactions}

{Usually, due to the absence of a fundamental theory regarding the
interactions in the dark sector, the functional form of }$Q${ is
introduced on a phenomenological basis in order to describe the interaction
rate between matter and scalar field. Following the literature, in our work we
utilize one of the most simple and popular parametrizations of the interaction
rate (\cite{boehmer,interaction1} and references therein), namely }%
$Q=3\alpha_{m}\rho_{m}H$ {. This particular expression, as we shall see
below, provides a well-defined set of dimensionless variables. Now concerning
parameter }$a_{m}${ the situation is as follows. For }$a_{m}>0${
the scalar field decays into matter and vice versa when }$a_{m}<0${.}

By substituting the latter expression into Eqs.(\ref{Eq2}) and (\ref{Eq3}) we
obtain
\begin{equation}
\rho_{m}=\rho_{m0}a^{-3(1+w_{m}-\alpha_{m})},\;\;\;\;\;\;\rho_{\phi}%
=\rho_{\phi0}g(a) \label{frie56}%
\end{equation}
with
\begin{equation}
\label{friga}g(a)=X(a)\left[  1-3\alpha_{m}\frac{\rho_{m0}}{\rho_{\phi0}}%
\int_{a}^{1}\frac{a^{-3(1+w_{\phi}-\alpha_{m})}}{X(a)}d\mathrm{ln}a\right]
\end{equation}
where $\frac{\rho_{m0}}{\rho_{\phi0}}=\frac{\Omega_{m0}}{\Omega_{\phi0}}$.
Following the methodology of the previous section we now provide the
normalized Hubble parameter:
\begin{equation}
E^{2}(a)=\Omega_{m0}a^{-3(1+w_{m}-\alpha_{m})}+\Omega_{\phi0}g(a)+2\Omega
_{Z_{t}}+2\sqrt{\Omega_{Z_{t}}}\sqrt{\Omega_{m0}a^{-3(1+w_{m}-\alpha_{m}%
)}+\Omega_{\phi0}g(a)+\Omega_{Z_{t}}}. \label{nffe22}%
\end{equation}
As expected in the case of $\alpha_{m}=0$, we get $g(a)=X(a)$ which implies
that the current model reduces to the minimally coupled FR cosmological model
[see Eq.(\ref{nfe22}].


\section{Dynamical Analysis}

\label{section3}

In this Section we use the method of critical points and the corresponding
eigenvalues of the gravitational field equations for a general potential, so
that we can study the various phases of the Finsler Randers gravity model and
to compare it with that of General Relativity. {Notice, that the
eigenvalues are important tools towards characterizing the stability of the
critical points. For example if a critical point is stable/attractor then the
corresponding eigenvalues have negative real parts. Therefore, the eigenvalues
can be used in order to understand the behavior of the dynamical system prior
to the critical point \cite{Liddle}.} Regarding the conditions of the cosmic
fluids involved in the analysis, the situation is as follows: (a) we consider
$Q=0$ which implies that the matter component is minimally coupled with the
scalar field, and (b) we allow interactions in the dark sector, namely
$Q=3\alpha_{m}\rho_{m}H$.

\subsection{Minimally coupled fluids}

The first step here is to introduce a new set of dimensionless variables
\cite{con1,con2}%
\begin{equation}
x=\frac{\dot{\phi}}{\sqrt{6}H}~,~y=\frac{\sqrt{V}}{\sqrt{3}H}~,~\Omega
_{z}=\frac{Z_{t}}{H}~,~\Omega_{m}=\frac{\rho_{m}}{3H^{2}}, \label{ss.01}%
\end{equation}
we have set $8\pi G=1$ and $c=1$. Therefore, the first Friedmann's equation
(\ref{e1}) becomes%
\begin{equation}
\left(  1+\Omega_{z}\right)  -x^{2}-y^{2}=\Omega_{m}, \label{ss.02}%
\end{equation}
or%

\begin{equation}
\left(  1+\Omega_{z}\right)  -\Omega_{\phi}=\Omega_{m},
\end{equation}
where $\Omega_{\phi}~$is the energy density of the scalar field, i.e.
$\Omega_{\phi}=x^{2}+y^{2}$.

Moreover, the field equations (\ref{e2}), (\ref{Eq1}), (\ref{Eq2}) and
(\ref{Eq3}) are given by the following system of first-order ordinary
differential equations
\begin{equation}
\frac{dx}{dN}=-3x+x(1+\frac{3}{4}\Omega_{Z}\ +\frac{1}{2}\Omega_{m}%
(1+3w_{m})+2x^{2}-y^{2})+\sqrt{\frac{3}{2}}\lambda y^{2}, \label{ss.03}%
\end{equation}

\begin{equation}
\frac{dy}{dN}=y(1+\frac{3}{4}\Omega_{Z}\ +\frac{1}{2}\Omega_{m}(1+3w_{m}%
)+2x^{2}-y^{2}-\sqrt{\frac{3}{2}\lambda x}), \label{ss.04}%
\end{equation}%
\begin{equation}
\frac{d\Omega_{z}}{dN}=\Omega_{Z}\left(  1+\frac{3}{4}\Omega_{Z}\ +\frac{1}%
{2}\Omega_{m}(1+3w_{m})+2x^{2}-y^{2}\right)  , \label{ss.05}%
\end{equation}
and%
\begin{equation}
\frac{d\lambda}{dN}=\sqrt{6}x\lambda^{2}[1-\Gamma\left(  \lambda\right)  ],
\label{ss.06}%
\end{equation}
with
\begin{equation}
\lambda=-\frac{V_{,\phi}}{V}~,~\Gamma=\frac{V_{,\phi\phi}V}{V_{,\phi}^{2}}.
\label{ss.07}%
\end{equation}
where $N=\ln\left(  a\right)  $ is the new lapse function, $V_{,\phi}%
=dV/d\phi$ and $V_{,\phi\phi}=d^{2}V/d\phi^{2}$. Moreover, we calculate that%
\begin{equation}
\frac{\dot{H}}{H^{2}}=-1-\frac{3}{4}\Omega_{z}-2x^{2}+y^{2}-\frac{1}{2}%
\Omega_{m}(1+3w_{m}), \label{ss.08}%
\end{equation}
hence the total equation of the state parameter $w_{\mathrm{tot}}$ can be
explicitly derived in terms of the new variables $x,~y$ and $\Omega_{z}$.
Indeed, after some calculations we arrive at%

\begin{equation}
w_{tot}=-1-\frac{2}{3}\frac{\dot{H}}{H^{2}}=- \frac{1}{3}+\frac{1}{2}%
\Omega_{z}+\frac{4}{3}x^{2}-\frac{2}{3}y^{2}+\frac{1}{3}\Omega_{m}(1+3w_{m}),
\label{ss09}%
\end{equation}

The dynamical system (\ref{ss.02})-(\ref{ss.06}) is an algebraic-differential
system which is valid for a general potential $V(\phi)$. Notice, that the
expression (\ref{ss.02}) is a constrain equation.
In this framework, we observe from (\ref{ss.02}) that the pair $\left(
x,y\right)  $ belongs to circle of radius $\sqrt{\left(  1+\Omega_{z}%
-\Omega_{m}\right)  }$.
Concerning the functional form of the potential energy we consider two cases,
namely exponential with $V_{1}\left(  \phi\right)  =V_{0}e^{-\lambda\phi}$ and
hyperbolic with $V_{2}\left(  \phi\right)  =V_{0}\cosh^{q}(p\phi)$. In the
exponential case the variable $\lambda$ is always constant and thus equation
(\ref{ss.06}) is satisfied identically, hence we can reduce the degrees of
freedom of the above system of equations. On the other hand it has been found
in the context of GR the hyperbolic potential provides a deSitter universe as
a late time attractor \cite{apBB}, while if we expand the hyperbolic form of
the potential as a Taylor series then we can approach the power-law and the
exponential potentials respectively.

\subsubsection{Exponential potential}

We continue our study by using the exponential potential.{ It is worth
noting that in this case the rhs of equation (\ref{ss.06}) is identical to
zero, which means that the resulting number of degrees of freedom is three.
The free variables are }$\left\{  x,y,\Omega_{z}\right\}  ${.
Concerning the number of degrees of freedom for the current potential we refer
the reader the work of \cite{Ame10}.}

The dynamical system (\ref{ss.03})-(\ref{ss.05}) contains six critical points,
among which five points are similar to those of GR and only one is a new
critical point accommodated by FR gravity model. Specifically, the
corresponding critical points are:%

\begin{table}[tbp] \centering
\caption{Critical points and cosmological parameters for scalar field cosmology in the Finsler-Randers theory for exponential potential.}%
\begin{tabular}
[c]{ccccccc}\hline\hline
\textbf{Point} & $(x,y,\Omega_{z})$ & $\mathbf{\Omega}_{m}$ & $\mathbf{\Omega
}_{z}$ & $\mathbf{w}_{tot}$ & \textbf{Acceleration} & \textbf{Existence}%
\\\hline
$P_{1}^{\pm}$ & $(\pm1,0,0)$ & $0$ & $0$ & $1$ & No & Always\\
$P_{2}$ & $(0,0,0)$ & $1$ & $0$ & $w_{m}$ & Yes for $w_{m}<-1/3$ & Always\\
$\mathbf{P}_{3}$ & $(\frac{\lambda}{\sqrt{6}},\mathbf{\pm}\sqrt{1-\frac
{\lambda^{2}}{6}},0)$ & $0$ & $0$ & \bigskip$-1+\frac{\lambda^{2}}{3}$ &
$-\sqrt{2}\leq\lambda\leq\sqrt{2}$ & $\lambda^{2}<6$\\
$\mathbf{P}_{4}$ & $(\sqrt{\frac{3}{2}}\frac{1+w_{m}}{\lambda},\mathbf{\pm
}\sqrt{\frac{3}{2}}\ \frac{\sqrt{1-w_{m}^{2}}}{\lambda},0)$ & $\frac
{\lambda^{2}-3(1+w_{m})}{\lambda^{2}}$ & $0$ & $w_{m}$ & Yes for $w_{m}<-1/3$
& $\frac{\lambda^{2}}{3}-1\geqq w_{m}>-1$\\
$F_{1}$ & $(0,0,-\frac{6(w_{m}+1)}{6w_{m}+5})$ & $1-\frac{6(w_{m}+1)}%
{6w_{m}+5}$ & $-\frac{6(w_{m}+1)}{6w_{m}+5}$ & $-1$ & Always & $w_{m}%
\neq-\frac{5}{6}$, $\rho_{m}~$violates S.E.C.\\\hline\hline
\end{tabular}
\label{exp1}%
\end{table}%

\begin{itemize}
\item Points $P_{1}^{\pm}$, with coordinates $(x,y,\Omega_{z})=(\pm1,0,0)~$
describe a phase of the universe which is dominated by the kinetic term of the
scalar field, where $\Omega_{m}=\Omega_{z}=0$ and $\Omega_{\phi}=1$ such that
$w_{\phi}=w_{\mathrm{tot}}=1$. Also, we compute the corresponding eigenvalues,
namely $m_{1}=3$, $m_{2}=\frac{1}{2}(6+\sqrt{6}\lambda)$ and $m_{3}%
=3(1-w_{m})$. Furthermore, these points are always unstable.

\item Point $P_{2}$ with coordinates $(x,y,\Omega_{z})=(0,0,0)$ describes the
matter dominated era of the universe, because $\Omega_{m}=1$, and $\Omega
_{z}=\Omega_{\phi}=0$ with $w_{\mathrm{tot}}=w_{m}.~$ Hence, for $w_{m}%
<-\frac{1}{3}$ we have an accelerating universe. Therefore, in this case
matter mimics dark energy. Notice, that point $P_{2}$ is always unstable.

\item Point $P_{3}$ with coordinates $(x,y,\Omega_{z})=(\frac{\lambda}%
{\sqrt{6}},\sqrt{1-\frac{\lambda^{2}}{6}},0)~$. At this point we find
$\Omega_{m}=\Omega_{z}=0$ and $\Omega_{\phi}=1$, hence the scalar filed is the
dominant component in the cosmic fluid while we have no presence of matter.
The equation of state parameter is given to be $w_{\mathrm{tot}}%
=-1+\frac{\lambda^{2}}{3}$. On the other hand point $P_{3}$ exists when
$\lambda^{2}<6$ (or $y\in%
\mathbb{R}
$). In the limit of $\lambda^{2}=6$ we recover points $P_{1}^{\pm}$. However,
for GR point $P_{3}$ is stable when $\lambda^{2}<3\left(  1+w_{m}\right)  $
and it is saddle when $3\left(  1+w_{m}\right)  <\lambda^{2}<6$. In the case
of FR model one of the eigenvalues, namely $\frac{\lambda^{2}}{2}$ is always
positive for real $\lambda$, so $P_{3}$ is always unstable.


\item Point $P_{4}$ with coordinates $(x,y,\Omega_{z})=\sqrt{\frac{3}{2}%
}(\frac{1+w_{m}}{\lambda},\ \frac{\sqrt{1-w_{m}^{2}}}{\lambda},0)$. This
critical point describes a tracker solution in which $\Omega_{z}=0$,
$\Omega_{m}=\frac{\lambda^{2}-3(1+w_{m})}{\lambda^{2}}$ and $w_{\phi
}=w_{\mathrm{tot}}=w_{m}$. Notice, that when $\frac{\lambda^{2}}{3}-1\geqq
w_{m}>-1$, point $P_{4}$ exists. In the case of FR gravity one of the
eigenvalues of the linearized system is always positive for $w_{m}>-1$, hence
point $P_{4}\left(  -1\right)  $ is always unstable, while the stability of
$P_{4}\left(  1\right)  $ holds for GR.

\item Point $F_{1}$ exists only for the FR model. The corresponding
coordinates and cosmological parameters are: $(x,y,\Omega_{z})=(0,0,-\frac
{6(w_{m}+1)}{6w_{m}+5})$, $\Omega_{z}=-\frac{6(w_{m}+1)}{6w_{m}+5}$,
$\Omega_{m}=1-\frac{6(w_{m}+1)}{6w_{m}+5}$ and $w_{\mathrm{tot}}=-1$. At point
$F_{1}$ we find that the contribution from the scalar field is negligible.
This point exists when $w_{m}\neq-\frac{5}{6}$, while for $w_{m}\in\left(
-1,1\right)  $, the current point exists only as long as $\rho_{m}$ violates
the strong energy condition. Lastly, point $P_{1}$ is always a stable
attractor when $w_{m}\in\left(  -1,1\right)  $ and it describes a de Sitter
point, because $w_{\mathrm{tot}}=-1.$
\end{itemize}

In Tables I and II, one may see a more compact presentation of the critical
points including coordinates, physical parameters, eigenvalues and
stability\footnote{In Table \ref{exp2}, $A=\sqrt{24\lambda^{2}-7\lambda
^{4}+24\lambda^{2}w_{m}-2\lambda^{4}w_{m}-24\lambda^{2}w_{m}^{2}+9\lambda
^{4}w_{m}^{2}-24\lambda^{2}w_{m}^{3}}$}. Now we proceed with the hyperbolic potential.%

\begin{table}[tbp] \centering
\caption{Eigenvalues and stability for the critical points of scalar field cosmology in the Finsler-Randers theory for exponential potential.}$%
\begin{tabular}
[c]{ccccc}\hline\hline
\textbf{Point/Eigenvalues} & $\mathbf{m}_{1}$ & $\mathbf{m}_{2}$ &
$\mathbf{m}_{3}$ & \textbf{Stable?}\\\hline
$P_{1}$ & $3$ & $\frac{1}{2}(6+\sqrt{6}\lambda)$ & $3(1-w_{m})$ & No\\
$P_{2}$ & $-\frac{3}{2}(1-w_{m})$ & $\frac{3}{2}(1+w_{m})$ & $\frac{3}%
{2}(1+w_{m})$ & No\\
$P_{3}$ & $\frac{\lambda^{2}}{2}$ & $\frac{1}{2}(-6+\lambda^{2})$ &
$-3+\lambda^{2}-3w_{m}$ & No\\
$P_{4}$ & $\frac{3}{2}(1+w_{m})$ & $-\frac{3}{4}(1-w_{m})-\frac{A}%
{4\lambda^{2}}$ & $-\frac{3}{4}(1-w_{m})+\frac{A}{4\lambda^{2}}$ & No\\
$F_{1}$ & $-3$ & $0$ & $-\frac{3}{2}(1+w_{m})$ & C.M.T.\ / Yes for $w_{m}%
>-1$\\\hline\hline
\end{tabular}
\ \ \ $\label{exp2}%
\end{table}%

\subsubsection{Hyperbolic potential}

Using the hyperbolic potential $V=V_{0}\cosh^{q}(p\phi)$ equation
(\ref{ss.06}) becomes%

\begin{equation}
\frac{d\lambda}{dN}=\frac{\sqrt{6}}{q}(qp-\lambda)(qp+\lambda)x\;.
\end{equation}
The critical points of the dynamical system are:

\begin{itemize}
\item Points $\bar{P}_{1}^{\pm}$ with coordinates $\left(  x,y,\lambda
,\Omega_{z}\right)  =(\pm1,0,\pm pq,0)$. Again at points $\bar{P}_{1}^{\pm}$
the universe is dominated by the kinetic term of the scalar field. These
points always exist and they are unstable. Notice that the critical points
$\bar{P}_{1}^{\pm}$ have the same dynamical properties with those of
$P_{1}^{\pm}$ (see exponential potential).

\item Point $\bar{P}_{2}$ with coordinates $\ \left(  x,y,\lambda,\Omega
_{z}\right)  =(0,0,\lambda_{c},0)$, where $\lambda_{c}$ is an arbitrary
parameter. At this point the universe is dominated by matter $\rho_{m}$, such
that, $\Omega_{m}=1$ and $w_{m}=1$. From the corresponding eigenvalues we find
that this critical point is always unstable.

\item Points $\bar{P}_{3}^{\pm}$ have coordinates $\left(  x,y,\lambda
,\Omega_{z}\right)  =\left(  \pm\frac{pq}{\sqrt{6}},\sqrt{1-\frac{(pq)^{2}}%
{6}},\pm pq,0\right)  $, and they are always unstable since their eigenvalues
are positive. The current critical point has the same dynamical properties
with that of $P_{3}$ (see exponential potential).

\item Points $\bar{P}_{4}^{\pm}$ have the coordinates of $P_{4}$ (exponential
potential) when $\lambda=pq$ which implies that the above points are always
unstable and they share the same dynamical properties (tracker solution).

\item Point $\bar{P}_{5}$ is the de Sitter solution, $\Omega_{m}=\Omega_{z}=0$
and $\Omega_{\phi}=1$, $w_{m}=-1$. The scalar field behaves as a cosmological
constant and the matter component vanished. It is interesting to mention that
the critical point $\bar{P}_{5}$ is stable and it could be the future
attractor of the universe when $p^{2}q<0$. On the other hand, if $p^{2}q>0$
then point $\bar{P}_{5}$ describes the past expansion era of the universe.

\item Point $\bar{F}_{1}$ exists only for FR model. The fact that $\lambda
_{c}$ is an arbitrary parameter implies that the current is exactly the same
with that of the exponential potential (see point $F_{1}$). Therefore, the
exponential and the hyperbolic potentials share the same properties.


\item Point $\bar{F}_{2}$ with coordinates $\left(  x,y,\lambda,\Omega
_{z}\right)  =(0,y_{c},0,-\frac{6(w_{m}+1+y^{2}+w_{m}y^{2})}{6w_{m}+5})$
appears only in the case of the FR model. This critical point exists for
$w_{m}\neq-\frac{5}{6}$, while when $w_{m}>\frac{5}{6}$ the point exists if
and only if $\rho_{m}$ violates the strong energy condition. The central
manifold theorem shows that point $\bar{F}_{2}$ is always unstable. Combining
the latter with the de Sitter solution ($w_{\mathrm{tot}}=-1$) we conclude
that the current critical point describes the past acceleration phase of the universe.
\end{itemize}

We collect our results in Tables III and IV\footnote{Where $B=3(-p^{2}%
q^{2}+p^{2}q^{2}w_{m}-$$\sqrt{p^{2}q^{2}(24-7p^{2}q^{2}+24w_{m}-2p^{2}%
q^{2}w_{m}-24w_{m}^{2}+9p^{2}q^{2}w_{m}^{2}-24w_{m}^{3})}),C=\frac
{\sqrt{3(3+4p^{2}qy^{2})}}{2}$}. However what it is interesting is that
various points which describe de Sitter phases of the universe exist in
contrary to the case of GR, if of course the matter source can violate the
strong energy condition.

%

\begin{table}[tbp] \centering
\caption{Critical points and cosmological parameters for scalar field cosmology in the Finsler-Randers theory for hyperbolic potential.}%
\begin{tabular}
[c]{cccccc}\hline\hline
\textbf{Point} & $(x,y,\lambda,\Omega_{z})$ & $\mathbf{\Omega}_{m}$ &
$\mathbf{\Omega}_{\phi}$ & $\mathbf{w}_{tot}$ & \textbf{Existence}\\\hline
$\bar{P}_{1}^{\pm}$ & $(\pm1,0,\pm pq,0)$ & $0$ & $1$ & $1$ & Always\\
$\bar{P}_{2}$ & $\ (0,0,\lambda_{c},0)$ & $1$ & $0$ & $-\frac{1}{3}+\frac
{1}{2}(1+3w_{m})$ & Always\\
$\bar{P}_{3}^{\pm}$ & $\left(  \pm\frac{pq}{\sqrt{6}},\sqrt{1-\frac{(pq)^{2}%
}{6}},\pm pq,0\right)  $ & $0$ & $1$ & $-1+\frac{(pq)^{2}}{3}$ & $(pq)^{2}%
\leq6$\\
$\bar{P}_{4}^{\pm}$ & $\left(  \pm\sqrt{\frac{3}{2}}\frac{1+w_{m}}{pq}%
,\ \sqrt{\frac{3}{2}}\frac{\sqrt{1-w_{m}^{2}}}{pq},\pm pq,0\right)  $ &
$\frac{(pq)^{2}-3(1+w_{m})}{(pq)^{2}}$ & $1-\Omega_{m}$ & $w_{m}$ &
$\frac{(pq)^{2}}{3}-1\geqq w_{m}>-1$\\
$\bar{P}_{5}$ & $(0,1,0,0)$ & $0$ & $1$ & $-1$ & Always\\
$\bar{F}_{1}$ & $\ (0,0,\lambda_{c},-\frac{6(w_{m}+1)}{6w_{m}+5})$ &
$1-\frac{6(w_{m}+1)}{6w_{m}+5}$ & $0$ & $-1$ & $w_{m}\neq-\frac{5}{6}$,
$\rho_{m}~$can$~$violate S.E.C.\\
$\mathbf{\bar{F}}_{2}$ & $\ (0,y_{c},0,\mathbf{-}\frac{6(w_{m}+1-y^{2}%
-w_{m}y^{2})}{6w_{m}+5})$ & $1-\Omega_{z}-y_{c}^{2}$ & $y_{c}^{2}$ & $-1$ &
$w_{m}\neq-\frac{5}{6}$,$~\rho_{m}~$can violate S.E.C.\\\hline\hline
\end{tabular}
\label{hyper1}%
\end{table}%
%

\begin{table}[tbp] \centering
\caption{Eigenvalues and stability for the critical points of scalar field cosmology in the Finsler-Randers theory for hyperbolic potential.}$%
\begin{tabular}
[c]{cccccc}\hline\hline
\textbf{Point/Eigenvalues} & $\mathbf{m}_{1}$ & $\mathbf{m}_{2}$ &
$\mathbf{m}_{3}$ & $\mathbf{m}_{4}$ & \textbf{Stable?}\\\hline
$\bar{P}_{1}^{\pm}$ & $3$ & $\pm2\sqrt{6}p$ & $\frac{1}{2}(6-\sqrt{6}pq)$ &
$\pm3(1-w_{m})$ & No\\
$\bar{P}_{2}$ & $0$ & $-\frac{3}{2}(1-w_{m})$ & $\frac{3}{2}(1+w_{m})$ &
$\frac{3}{2}(1+w_{m})$ & No\\
$\bar{P}_{3}^{\pm}$ & $-2p^{2}q$ & $\frac{p^{2}q^{2}}{2}$ & $\frac{1}%
{2}(-1+p^{2}q^{2})$ & $-3+p^{2}q^{2}-3w_{m}$ & No\\
$\bar{P}_{4}^{\pm}$ & $\frac{-6(1+w_{m})}{q}$ & $\frac{3(1+w_{m})}{2}$ &
$-\frac{3}{4}(1-w_{m})-\frac{B}{4p^{2}q^{2}}$ & $-\frac{3}{4}(1-w_{m}%
)+\frac{B}{4p^{2}q^{2}}$ & No\\
$\bar{P}_{5}$ & $0$ & $-\frac{1}{2}(3+\sqrt{9+12p^{2}q})$ & $-\frac{1}%
{2}(3-\sqrt{9+12p^{2}q})$ & $-3(1+w_{m})$ & Yes for $p^{2}q<0$\\
$\bar{F}_{1}$ & $0$ & $0$ & $-3$ & $-\frac{3}{2}(1+w_{m})$ & No\\
$\bar{F}_{2}$ & $0$ & $-\frac{3}{2}(1+w_{m})(1+y^{2})$ & $-\frac{3}{2}%
+\frac{C}{2(5+6w_{m})}$ & $-\frac{3}{2}-\frac{C}{2(5+6w_{m})}$ &
No\\\hline\hline
\end{tabular}
\ $\label{hyper2}%
\end{table}%

\subsection{\textbf{Interaction of Scalar field and matter}}

Here we allow interactions between scalar field and matter for which the
interaction rate is given by $Q=a_{m}\rho_{m}H$ ~\cite{interaction1,boehmer}.
In this context
the gravitational field equations (\ref{e1}),~ (\ref{e2}), (\ref{Eq1}),
(\ref{Eq2}) and (\ref{Eq3}) \ are described by the system (\ref{ss.02}%
)-(\ref{ss.07}) where now equation (\ref{ss.03}) becomes%
\begin{equation}
\frac{dx}{dN}=-3x+x(1+\frac{3}{4}\Omega_{Z}\ +\frac{1}{2}\Omega_{m}%
(1+3w_{m})+2x^{2}-y^{2})+\sqrt{\frac{3}{2}}\lambda y^{2}-\bar{Q}%
\end{equation}
\qquad in which we have set $\bar{Q}=\frac{Q}{6H^{3}x}$. In the case of
$Q=a_{m}\rho_{m}H$ it is easy to show $\bar{Q}=\frac{\alpha_{m}}{2}\Omega
_{m}.~\ $ It is interesting to mention that if $Q>0$ then the scalar field
decays into matter, while the opposite holds for $Q<0$. In contrast to section
3.1 here we verify that for $\lambda^{2}=pq$ the two potentials share the same
critical points. Therefore, we focus our analysis on the exponential case.



Specifically, we find the following family of critical points:


\begin{itemize}
\item Points ~$A_{1}^{\pm}$ with coordinates~$(x,y,\Omega_{z})=(\pm1,0,0)$. At
these points the universe is dominated by the kinetic term of the scalar
field. These points are always unstable.

\item Points~$A_{2}^{\pm}$ have coordinates $\ (x,y,\Omega_{z})=\left(
\pm\sqrt{\frac{a_{m}}{3(1-w_{m})}},0,0\right)  $ and $\Omega_{m}%
=1-\Omega_{\phi}=1-\frac{a_{m}}{3(1-w_{m})}$ It is easy to check that when
$a_{m}\rightarrow0$ points $A_{2}^{\pm}\ $ reduce to $P_{2}$ (see Table
\ref{exp1}). The current points exist only as long as $a_{m}>0$, where we have
assumed $\left\vert w_{m}\right\vert <1$.
Here, only the kinetic term of the scalar field survives which means that
$w_{\phi}=1$, while the total equation of the state parameter is
$w_{\mathrm{tot}}=w_{m}+\frac{a_{m}}{3}$. These points are always unstable.

\item Point $A_{3}$ with coordinates $\ (x,y,\Omega_{z})=\left(  \frac
{\lambda}{\sqrt{6}},\sqrt{1-\frac{\lambda^{2}}{6}},0\right)  $. Practically,
this point coincides with point~$P_{3}~$of Table \ref{exp1}, where only the
scalar field participates in the dynamics. The current point is always
unstable and exists for~$\lambda^{2}\leq6$.

\item Point $A_{4}$ can be viewed as a generalization of point $P_{4}$ (see
Table~\ref{exp1}), where now the corresponding coordinates are $(x,y,\Omega
_{z})=\left(  \frac{3(1+w_{m})+a_{m}}{\sqrt{6}\lambda},y\left(  A_{4}\right)
,0\right)  $, where
\[
y\left(  A_{4}\right)  =\sqrt{\frac{9+6a_{m}+a_{m}^{2}-2a_{m}\lambda
^{2}+9w_{m}-a_{m}^{2}w_{m}-9w_{m}^{2}-6a_{m}w_{m}^{2}-9w_{m}^{3}}{6\lambda
^{2}(1+w_{m})}.}%
\]
Point $A_{4}$ exists when $\lambda^{2}\geq\ 3(1+w_{m})+a_{m}$ and it describes
an accelerated universe for $w_{m}+\frac{a_{m}}{3}<-1/3$. This point is always unstable.

\item Finally, point~$R_{1}$ appears only in the case of the FR model and
practically it extents $F_{1}$ (see Table~\ref{exp1}) for the minimally
coupled fluids. The coordinates of the point are $(x,y,\Omega_{z})=\left(
\ -\sqrt{\frac{-a_{m}}{30+11a_{m}+36w_{m}}},0,-\frac{12(3w_{m}+3+a_{m}%
)}{30+11a_{m}+36w_{m}}\right)  $. In contrast, to point $F_{1}$ the dynamical
situation of $R_{1}$ is strongly affected by the parameter $a_{m}$ which is
related with the interaction rate $Q$.
Specifically, $R_{1}$ exists when $\frac{-a_{m}}{30+11a_{m}+36w_{m}}>0$, and
$\Omega_{m}\,\ $ is not violating the strong energy condition when
$30+11a_{m}+36w_{m}<0$. Hence if $a_{m}<0$, \ then ~for $w_{m}<\frac
{-30+11\left\vert a_{m}\right\vert }{36}$ \ the strong energy condition holds
and for $\left\vert w_{m}\right\vert <1$, we find that $a_{m}\in\left(
-\frac{6}{11},0\right)  $. From the eigenvalues of the linearized system close
to the critical point we find that it is stable when$~-6<a_{m}<0~$%
and$~7a_{m}+\sqrt{9-174a_{m}+313\left(  a_{m}\right)  ^{2}}<w_{m}$ or when
$0<a_{m}<\frac{87-12\sqrt{33}}{313},~$and$~$%
\[
-33+7a_{m}-\sqrt{9-174a_{m}+313\left(  a_{m}\right)  ^{2}}<36w_{m}%
<-33+7a_{m}+\sqrt{9-174a_{m}+313\left(  a_{m}\right)  ^{2}}.
\]
Furthermore, in the limit in which $w_{m}=0$, we find that the point $R_{1}$
is stable when $-\frac{3\left(  2+\sqrt{59}\right)  }{11}<a_{m}<0.$
\end{itemize}

We summarize our results in Tables \ref{inExp1} and \ref{inExp2}.%

\begin{table}[tbp] \centering
\caption{Critical points and cosmological parameters for scalar field cosmology in the Finsler-Randers theory with interaction for exponential potential.}%
\begin{tabular}
[c]{ccccc}\hline\hline
\textbf{Point} & $\mathbf{(x,y,\Omega}_{z}\mathbf{)}$ & $\mathbf{\Omega}_{m}$
& $\mathbf{\Omega}_{z}$ & $\mathbf{w}_{tot}$\\\hline
$A_{1}^{\pm}$ & $(\pm1,0,0)$ & $0$ & $0$ & $1$\\
$A_{2}^{\pm}$ & $\ (\pm\sqrt{\frac{a_{m}}{3(1-w_{m})}},0,0)$ & $1-\frac{a_{m}%
}{3(1-w_{m})}$ & $0$ & $w_{m}+\frac{a_{m}}{3}$\\
$A_{3}$ & $\ (\frac{\lambda}{\sqrt{6}},\sqrt{1-\frac{\lambda^{2}}{6}},0)$ &
$0$ & $0$ & $-1+\frac{\lambda^{2}}{3}$\\
$A_{4}$ & $\ (\frac{3(1+w_{m})+a_{m}}{\sqrt{6}\lambda},y\left(  A_{4}\right)
,0)$ & $\frac{(3(1+w_{m})+a_{m})^{2}(\frac{\lambda^{2}}{(3(1+w_{m})+a_{m}%
)}-1)}{3\lambda^{2}(1+w_{m})}$ & $0$ & $w_{m}+\frac{a_{m}}{3}$\\
$R_{1}$ & $\ (-\sqrt{\frac{-a_{m}}{30+11a_{m}+36w_{m}}},0,-\frac
{12(3w_{m}+3+a_{m})}{30+11a_{m}+36w_{m}})$ & $-\frac{6}{30+11a_{m}+36w_{m}}$ &
$-\frac{12(3w_{m}+3+a_{m})}{30+11a_{m}+36w_{m}}$ & $-1$\\\hline\hline
\end{tabular}
\label{inExp1}%
\end{table}%
%

\begin{table}[tbp] \centering
\caption{Critical and stability for scalar field cosmology with interaction in the Finsler-Randers theory for exponential potential.}%
\begin{tabular}
[c]{cccc}\hline\hline
\textbf{Point} & \textbf{Acceleration} & \textbf{Existence} &
\textbf{Stability}\\\hline
$A_{1}$ & No & yes & No\\
$A_{2}$ & Yes for $w_{m}+\frac{a_{m}}{3}<-1/3$ & $0<\frac{a_{m}}{3(1-w_{m}%
)}<1$ & No\\
$A_{3}$ & Yes$~-\sqrt{2}\leq\lambda\leq\sqrt{2}$ & $\lambda^{2}\leq6$ & No\\
$A_{4}$ & Yes for $w_{m}+\frac{a_{m}}{3}<-1/3$ & $\lambda^{2}\geq
\ 3(1+w_{m})+a_{m}$ & No\\
$R_{1}$ & Yes & Yes for $~\frac{-a_{m}}{30+11a_{m}+36w_{m}}>0$ & Yes for
specific $a_{m},~w_{m}$\\\hline\hline
\end{tabular}
\label{inExp2}%
\end{table}%

\section{Linear growth of matter Perturbations}

\label{section4} In this section we study the linear growth of matter
fluctuations within the context of FR cosmology. We will then compare our
results with those of the DGP and $\Lambda$CDM models. This can help us to
appreciate the relative differences and similarities of the above cosmological
models at the perturbation level. Owing to the fact that we are well inside in
the matter epoch we can neglect the radiation component from the Hubble expansion.

Let us start, with the basic differential equation which provides the
evolution of linear matter perturbations
\cite{Lue04,Linder2007,Gann09,Stab06,Uzan07,Tsu08,Steigerwald:2014ava}
\begin{equation}
\ddot{\delta}_{m}+2\nu H\dot{\delta}_{m}-4\pi G\mu\rho_{m}\delta_{m}=0\;.
\label{eq:111}%
\end{equation}
Notice, that a general solution of the above equation is given by $\delta
_{m}\propto D(t)$, where $D(t)$ is the linear growth factor usually scaled to
unity at the present epoch. As is well known, the quantities $\nu\equiv1+Q/H$
and $\mu\equiv G_{\mathrm{eff}}/G_{N}$ are related with the physics of dark
energy. Indeed, for scalar field dark energy models which are inside general
relativity $\nu$ and $\mu$ are both equal to unity. In the case of either
modified gravity models or inhomogeneous dark energy models (inside GR) we get
$\nu=1$ and $\mu\neq1$. Lastly, if we have interactions between dark matter
and dark energy then we get $\nu\neq1$ and $\mu\neq1$. In the current section
we do not allow interactions in the dark sector, hence $Q=0$ and $\nu=1$.
{Also, in our case one would expect to have $\mu\equiv G_{\mathrm{eff}}%
/G_{N}=1$. Recall that Stavrinos \& Diakogiannis \cite{18} showed that the
Ricci tensor in Finsler Randers gravity is written as the sum of the nominal
Ricci tensor plus a small tensor perturbation. Under these conditions, if we
utilize the linear perturbation theory in the gauge-invariant regime then we
can obtain Eq.(\ref{eq:111}) with $\nu=1$ and $\mu=G_{\mathrm{eff}}/G$=1.
Notice, that the full perturbation analysis in Finsler geometries are in
progress and will be published elsewhere.}

Another crucial parameter in linear growth studies is the growth rate of
clustering \cite{Peeb93}
\begin{equation}
f(a)=\frac{d\mathrm{ln}\delta_{m}}{d\mathrm{ln}a}\simeq\Omega_{m}^{\gamma
}(a)\;, \label{fzz221}%
\end{equation}
with
\begin{equation}
\Omega_{m}(a)=\frac{\Omega_{m0}a^{-3(1+w_{m})}}{E^{2}(a)} \label{ddomm}%
\end{equation}
where $E(a)=H(a)/H_{0}$ is the normalized Hubble parameter and $\gamma$ is the
growth index. From Eq.(\ref{ddomm}) we can obtain
\begin{equation}
\frac{d\Omega_{m}}{da}=-3\frac{\Omega_{m}(a)}{a}\left(  1+w_{m}+\frac{2}%
{3}\frac{d\mathrm{ln}E}{d\mathrm{ln}a}\right)  \;. \label{ddomm1}%
\end{equation}
Changing the variables from $t$ to $a$ ($\frac{d}{dt}=H\frac{d}{d\ln a}$) in
Eq.(\ref{eq:111}) and using Eqs.(\ref{fzz221},\ref{ddomm1}) we arrive at
\begin{equation}
\frac{df}{d\mathrm{ln}a}+\left(  2\nu+\frac{d{\ln}E}{d\mathrm{ln}a}\right)
f+f^{2}=\frac{3\mu\Omega_{m}}{2}. \label{fzz444}%
\end{equation}
Similar to \cite{Steigerwald:2014ava} we write Eq.(\ref{fzz444}) as
\begin{equation}
\frac{d\omega}{d{\ln}a}(\gamma+\omega\frac{d\gamma}{d\omega})+\mathrm{e}%
^{\omega\gamma}+2\nu+\frac{d{\ln}E}{d\mathrm{ln}a}=\frac{3}{2}\mu
\mathrm{e}^{\omega(1-\gamma)},
\end{equation}
where $\omega=\mathrm{ln}\Omega_{m}(a)$ which means that at $z\gg1$
($a\rightarrow0$) we have $\Omega_{m}(a)\rightarrow1$ [or $\omega\rightarrow
0$]. Regarding the evolution of growth index we use the methodology of
Steigerwald et al. \cite{Steigerwald:2014ava} (see also the relevant
discussion in \cite{Basola}), who proposed the following parametrization
\begin{equation}
\gamma(a)=\gamma_{0}+\gamma_{1}\omega(a).
\end{equation}
Obviously, under the latter parametrization the asymptotic value of the growth
index is $\gamma_{\infty}\approx\gamma_{0}$.

Notice, that within the theoretical treatment of \cite{Steigerwald:2014ava}
the coefficients $\gamma_{0}$ and $\gamma_{1}$ are given by
\begin{equation}
\gamma_{0}=\frac{3(M_{0}+M_{1})-2(\mathcal{H}_{1}+N_{1})}{2+2X_{1}+3M_{0}}
\label{g000}%
\end{equation}
and
\begin{align}
&  \gamma_{1}=3\frac{M_{2}+2M_{1}B_{1}(1-y_{1})+M_{0}B_{2}(1-y_{1},-y_{2}%
)}{2(2+4X_{1}+3M_{0})}\;\nonumber\label{g111}\\
&  -2\frac{B_{2}(y_{1},y_{2})+X_{2}\gamma_{0}+\mathcal{H}_{2}+N_{2}%
}{2(2+4X_{1}+3M_{0})}\,.
\end{align}
The following quantities have been{ defined as}:
\begin{equation}
X_{n}=\left.  \frac{d^{n}(d\omega/d\mathrm{ln}a)}{d\omega^{n}}\right\vert
_{\omega=0}\,\ \ M_{n}=\left.  \frac{d^{n}\mu}{d\omega^{n}}\right\vert
_{\omega=0} \label{Coef1}%
\end{equation}
and
\begin{equation}
N_{n}=\left.  \frac{d^{n}\nu}{d\omega^{n}}\right\vert _{\omega=0}%
,\label{Coef2}\\
\ \ \mathcal{H}_{n}=-\frac{1}{2}X_{n}=\left.  \frac{d^{n}(d\mathrm{ln}%
E/d\mathrm{ln}a)}{d\omega^{n}}\right\vert _{\omega=0}\,,
\end{equation}
with $\frac{d^{0}}{d\omega^{0}}\equiv1$. Here the absence of interactions in
the dark sector imposes $\nu=1$ which implies $N_{0}=1$ and $N_{n}=0$ for
$n\geq1$. In this context, $B_{1}$ and $B_{2}$ are the Bell polynomials of
first and second kind, namely $B_{1}(y_{1})=y_{1}$ and $B_{2}(y_{1}%
,y_{2})=y_{1}^{2}+y_{2}$. Based on Steigerwald et al.
\cite{Steigerwald:2014ava} [see their Eq.(10)] the pair $(y_{1},y_{2})$ is
equal to $(\gamma_{0},0)$ which provides $B_{1}(1-y_{1})=1-\gamma_{0}$,
$B_{2}(y_{1},y_{2})=\gamma_{0}^{2}$ and $B_{2}(1-y_{1},-y_{2})=(\gamma
_{0}-1)^{2}$. At this point we are ready to calculate the aforementioned
quantities in order to provide $\gamma_{0}$ and $\gamma_{1}$. Bellow we
present two different cases.


\subsubsection{Finsler Randers with Cold Dark Matter}

In this case we consider Cold Dark Matter (CDM), hence the corresponding
equation of state parameter is given by $p_{m}=0$ ($w_{m}=0$). Moreover, in
order to simplify the calculations we also neglect the scalar field from the
analysis $\rho_{\phi}(a)=0$. Under the latter conditions the quantities $\mu$
and $\frac{d\mathrm{ln}E}{d\mathrm{ln}a}$ become (see also \cite{Lue04,Gong10}%
)
\begin{equation}
\mu\equiv\frac{G_{\mathrm{eff}}(a)}{G_{N}}=\left\{
\begin{array}
[c]{cc}%
1 & \mbox{$\Lambda$CDM or FR}\\
\frac{2+4\Omega_{m}^{2}(a)}{3+3\Omega_{m}^{2}(a)}=\frac{2+4\mathrm{e}%
^{2\omega}}{3+3\mathrm{e}^{2\omega}} & \mbox{DGP.}
\end{array}
\right.  \label{OF11}%
\end{equation}
and
\begin{equation}
\frac{d\mathrm{ln}E}{d\mathrm{ln}a}=\left\{
\begin{array}
[c]{cc}%
-\frac{3\Omega_{m}(a)}{1+\Omega_{m}(a)}=-\frac{3\mathrm{e}^{\omega}%
}{1+\mathrm{e}^{\omega}} & \;\;\mbox{DGP or FR}\\
-\frac{3}{2}\Omega_{m}(a)=-\frac{3}{2}\mathrm{e}^{\omega} &
\mbox{$\Lambda$CDM}
\end{array}
\right.  \label{F11}%
\end{equation}
where for comparison we have included the $\Lambda$CDM and the DGP models
respectively. It is worth noting that in order to derive the last equalities
in Eqs.(\ref{OF11}-\ref{F11}) we have inverted the transformation of
Steigerwald et al. \cite{Steigerwald:2014ava}, namely $\Omega_{m}%
=\mathrm{e}^{\omega}$. Lastly, utilizing Eqs.(\ref{ddomm1}) and (\ref{F11})
the function $\frac{d\omega}{d\mathrm{ln}a}$ is written as
\begin{equation}
\frac{d\omega}{d\mathrm{ln}a}=\frac{d\mathrm{ln}\Omega_{m}}{d\mathrm{ln}%
a}=\left\{
\begin{array}
[c]{cc}%
-\frac{3\left[  1-\Omega_{m}(a)\right]  }{1+\Omega_{m}(a)}=-\frac{3\left(
1-\mathrm{e}^{\omega}\right)  }{1+\mathrm{e}^{\omega}} &
\;\;\mbox{DGP or FR}\\
-3\left[  1-\Omega_{m}(a)\right]  =-3\left(  1-\mathrm{e}^{\omega}\right)  . &
\mbox{$\Lambda$CDM}
\end{array}
\right.  \label{G11}%
\end{equation}

Under the aforementioned conditions we can provide for the first time the
growth coefficients of the FR cosmological model. In particular, we find the
following results:

\begin{itemize}
\item FR model. In this case we find
\[
\{ M_{0},M_{1},M_{2},N_{1},N_{2}\}=\{ 1,0,0,0,0\}
\]
\[
\{ \mathcal{H}_{1},\mathcal{H}_{2},X_{1},X_{2}\}=\{ -\frac{3}{4},0,\frac{3}%
{2},0\}
\]
and thus
\[
\gamma_{\infty}^{(FR)}\approx\gamma_{0}^{(FR)}=\frac{9}{16},\;\;\;\;\;\;\gamma
_{1}^{(FR)}=-\frac{15}{5632}\approx-0.0027.
\]

\item DGP model (see also \cite{Steigerwald:2014ava}). In this case we obtain
\[
\{ M_{0},M_{1},M_{2},N_{1},N_{2}\}=\{ 1,\frac{1}{3},0,0,0\}
\]
\[
\{ \mathcal{H}_{1},\mathcal{H}_{2},X_{1},X_{2}\}=\{ -\frac{3}{4},0,\frac{3}%
{2},0\}
\]
and thus
\[
\gamma_{\infty}^{(DGP)}\approx\gamma_{0}^{(DGP)}=\frac{11}{16}%
,\;\;\;\;\;\;\gamma_{1}^{(DGP)}=-\frac{7}{5632}\approx-0.0012.
\]

\item $\Lambda$CDM model (see also \cite{Steigerwald:2014ava}). In this case
we find
\[
\{M_{0},M_{1},M_{2},N_{1},N_{2}\}=\{1,0,0,0,0\}
\]%
\[
\{\mathcal{H}_{1},\mathcal{H}_{2},X_{1},X_{2}\}=\{-\frac{3}{2},-\frac{3}%
{2},3,3\}
\]
and thus
\[
\gamma_{\infty}^{(\Lambda)}\approx\gamma_{0}^{(\Lambda)}=\frac{6}%
{11},\;\;\;\;\;\;\gamma_{1}^{(\Lambda)}=-\frac{15}{2057}\approx-0.0073.
\]

\end{itemize}

\subsubsection{Finsler Randers with collissional matter}

Here we consider that matter has pressure $w_{m}=\frac{p_{m}}{\rho_{m}}\neq0$,
hence the matter density evolves differently from the usual power law
($\propto a^{-3}$), namely $\rho_{m}\propto a^{-3(1+w_{m})}$. In this case the
form of $\mu$ is given by $\mu=(1+3w_{m})(1+w_{m})$ \cite{Lima}, while again
for the FR model we utilize $\rho_{\phi}=0$. In this framework, the normalized
Hubble parameter becomes [see Eq.(\ref{EE11})]%

\begin{equation}
E(a)=\left\{
\begin{array}
[c]{cc}%
\sqrt{\Omega_{m0}a^{-3(1+w_{m})}+\Omega_{Z_{t}}}+\sqrt{\Omega_{Z_{t}}} &
\mbox{FR}\\
\sqrt{\Omega_{m0}a^{-3(1+w_{m})}+1-\Omega_{m0}} & \mbox{$\Lambda$-cosmology}
\end{array}
\right.  \label{EF11}%
\end{equation}
from which we derive
\begin{equation}
\frac{d\mathrm{ln}E}{d\mathrm{ln}a}=\left\{
\begin{array}
[c]{cc}%
-\frac{3(1+w_{m})\Omega_{m}(a)}{1+\Omega_{m}(a)}=-\frac{3(1+w_{m}%
)\mathrm{e}^{\omega}}{1+\mathrm{e}^{\omega}} & \;\;\mbox{FR}\\
-\frac{3}{2}(1+w_{m})\Omega_{m}(a)=-\frac{3}{2}(1+w_{m})\mathrm{e}^{\omega} &
\mbox{$\Lambda$-cosmology}
\end{array}
\right.  \label{F12}%
\end{equation}
and
\begin{equation}
\frac{d\omega}{d\mathrm{ln}a}=\frac{d\mathrm{ln}\Omega_{m}}{d\mathrm{ln}%
a}=\left\{
\begin{array}
[c]{cc}%
-\frac{3(1+w_{m})\left[  1-\Omega_{m}(a)\right]  }{1+\Omega_{m}(a)}%
=-\frac{3(1+w_{m})\left(  1-\mathrm{e}^{\omega}\right)  }{1+\mathrm{e}%
^{\omega}} & \;\;\mbox{FR}\\
-3(1+w_{m})\left[  1-\Omega_{m}(a)\right]  =-3(1+w_{m})\left(  1-\mathrm{e}%
^{\omega}\right)  & \mbox{$\Lambda$-cosmology.}
\end{array}
\right.  \label{G12}%
\end{equation}
Then based on Eqs. (\ref{Coef1}), (\ref{Coef2}), (\ref{F12}) and (\ref{G12}),
we find the following results:

\begin{itemize}
\item FR model. Here we obtain
\[
\{M_{0},M_{1},M_{2},N_{1},N_{2}\}=\{(1+3w_{m})(1+w_{m}),0,0,0,0\}
\]%
\[
\{\mathcal{H}_{1},\mathcal{H}_{2},X_{1},X_{2}\}=\{-\frac{3}{4}(1+w_{m}%
),0,\frac{3}{2}(1+w_{m}),0\},
\]
hence
\[
\gamma_{\infty}^{(FR)}\approx\gamma_{0}^{(FR\phi)}=\frac{9(1+w_{m})(1+2w_{m}%
)}{2(8+3w_{m}(5+3w_{m}))}%
\]
and
\[
\gamma_{1}^{(FR)}=-\frac{3(1+w_{m})(5+21w_{m})(1+3w_{m}(4+3w_{m}%
))}{8(11+9w_{m}(2+w_{m}))(8+3w_{m}(5+3w_{m}))^{2}}.
\]

\item $\Lambda$ cosmology. In this case we find
\[
\{M_{0},M_{1},M_{2},N_{1},N_{2}\}=\{(1+3w_{m})(1+w_{m}),0,0,0,0\}
\]%
\[
\{\mathcal{H}_{1},\mathcal{H}_{2},X_{1},X_{2}\}=\{-\frac{3}{2}(1+w_{m}%
),-\frac{3}{2}(1+3w_{m}),3(1+w_{m}),3(1+w_{m})\}
\]
and thus
\[
\gamma_{\infty}^{(\Lambda)}\approx\gamma_{0}^{(\Lambda)}= \frac{3(1+w_{m}%
)(2+3w_{m})}{11+9w_{m}(2+w_{m})}%
\]
and
\[
\gamma_{1}^{(\Lambda)}=-\frac{3(1+w_{m})(2+3w_{m})(5+3w_{m})(1+3w_{m}%
(4+3w_{m}))}{2(11+9w_{m}(2+w_{m}))^{2}(17+3w_{m}(8+3w_{m}))}.
\]

\end{itemize}

Finally, it is easy to show that for $w_{m}=0$ the above results reduce to
those of section 4.0.1 as they should.%

\[
\gamma_{\infty}^{(\Lambda)}\approx\gamma_{0}^{(\Lambda)}=\frac{6}%
{11},\;\;\;\;\;\;\gamma_{1}^{(\Lambda)}=-\frac{15}{2057}\approx-0.0073.
\]

\subsubsection{Finsler Randers with collissional matter and scalar field}

{Here we assume that the equation of state parameter of the scalar
field is} $w_{\phi}=\frac{p_{\phi}}{\rho_{\phi}}=const.$ {In this case
we have} $\rho_{\phi}=\rho_{\phi0}a^{-3(1+w_{\phi})}$. {Following the
aforementioned lines we calculate once more all the previous
derivatives\footnote{With the aid of Eq.(\ref{friga}) we obtain
\[
ag^{\prime}(a)=-3(1+w_{\phi})g(a)+3a_{m}\frac{\rho_{m0}}{\rho_{\phi0}%
}a^{-3(1+w_{\phi}-a_{m})},
\]
where the prime denotes to derivative with respect to the scale factor.} but
in this case we have to take into account the extra terms that are added due
to the non-zero derivative} $\frac{d\Omega_{\phi}}{d\ln a}$. {After
some algebra the main coefficients become}
\[
M_{0}=(1+3w_{m})(1+w_{m}),~M_{1,2}=0,~N_{1,2}=0,\mathcal{H}_{2}=X_{2}=0,
\]
{and}
\[
\mathcal{H}_{1}=\frac{1}{2}X_{1}=-\frac{3}{4}(1+w_{m})+\frac{9c_{\phi}}%
{4}(1-w_{m}+2w_{\phi})(2+w_{m}+w_{\phi}),
\]
{where} $c_{\phi}=lim_{a\rightarrow0}\left[  \frac{\Omega_{\phi}%
(a)}{(d\mathrm{ln}\Omega_{m}/d\mathrm{ln}a)}\right]  $.

{Inserting the above coefficients into Eqs.(\ref{g000}-\ref{g111}) we
can trivially calculate }$\left\{  \gamma_{0},\gamma_{1}\right\}  $%

\[
\gamma_{0}^{(FR\phi)}=\frac{9[c_{\phi}(1-w_{m}+2w_{\phi})(2+w_{m}+w_{\phi
})-(1+w_{m})(1+2w_{m})]}{2[c_{\phi}(1-w_{m}+2w_{\phi})(2+w_{m}+w_{\phi
})-8-3w_{m}(5+3w_{m})]}%
\]
{and}
\[
\gamma_{1}^{(FR\phi)}=-\frac{3[(1+w_{m})(1+3w_{m})(7+3w_{m}-9C)^{2}%
-54(C-(1+w_{m})(1+2w_{m}))^{2}]}{8(11+9w_{m}(2+w_{m})-18C)(8-9C+3w_{m}%
(5+3w_{m}))^{2}}.
\]

{where we have considered} $C=c_{\phi}(1-w_{m}+2w_{\phi})(2+w_{m}%
+w_{\phi}).$ {To conclude, it is easy to show that for} $c_{\phi}=0$
{the current results boils down to those of section 4.0.2 as they
should.}

\section{\textbf{Conclusions}}

{\label{conc}}

In this work we investigated the cosmological behavior of the Finsler-Randers
(FR) cosmological model at the background and at the perturbation level. First
we have provided the field equations in homogeneous and isotropic FR universe,
for which we assumed that the cosmic fluid contains matter, radiation and a
scalar field. Then for different types of FR models we performed a critical
point analysis in order to study the various phases of the FR gravity theory
and we compare the current results with those of General Relativity. For all
FR models we found solutions which accommodate cosmic acceleration and under
specific conditions they provide de-Sitter points as stable late-time attractors.

Second, we analytically studied for the first time the growth of perturbations
in FR cosmologies. Initially, for simplicity we neglected the scalar field
from the perturbation analysis. In the context of cold dark matter we found
that the asymptotic value of the growth index is $\gamma_{\infty}%
^{(FR)}\approx\frac{9}{16}$ which is somewhat greater ($\sim3\%$) than that of
$\Lambda$CDM model $\gamma_{\infty}^{(\Lambda)}\approx\frac{6}{11}$. Moreover,
it is worth noting that we confirmed the results of \cite{Bastav13}, namely
the FR model mimics the Dvali, Gabadadze and Porrati gravity model as far as
the cosmic expansion is concerned, while the two models can be distinguished
at the perturbation level, since the fractional difference $\Delta
\gamma_{\infty}(\%)=100\times\lbrack\gamma_{\infty}^{(FR)}-\gamma_{\infty
}^{(DGP)}]/\gamma_{\infty}^{(DGP)}$ is $\sim-18.2\%$, where $\gamma_{\infty
}^{(DGP)}\approx\frac{11}{16}$. On the other hand, if we allow pressure in the
matter fluid then we derived $\gamma_{\infty}^{(FR)}\approx\frac
{9(1+w_{m})(1+2w_{m})}{2[8+3w_{m}(5+3w_{m})]}$ and $\gamma_{\infty}%
^{(\Lambda)}\approx\frac{3(1+w_{m})(2+3w_{m})}{11+9w_{m}(2+w_{m})}$, where
$w_{m}=p_{m}/\rho_{m}$ is the matter equation of state parameter. Finally, we
generalized the growth index analysis by including the effects of the scalar
field and we found that the evolution of the growth index in FR cosmologies is
affected by the scalar field.

\begin{acknowledgments}
SB acknowledges support by the Research Center for Astronomy of the Academy of
Athens in the context of the program ''Testing general relativity on
cosmological scales'' (ref. number 200/872). AP acknowledges the financial
support of FONDECYT grant no. 3160121 and thanks the Durban University of
Technology for the hospitality provided while part of this work was performed.
\end{acknowledgments}

\end{document}